\documentclass[twocolumn]{aastex631}

\usepackage{xspace}

\usepackage{amsmath}
\usepackage{natbib}
\usepackage{enumitem}
\usepackage{caption}
\usepackage{graphicx}
\usepackage[T1]{fontenc}
\usepackage{booktabs}
\usepackage[encapsulated]{CJK}
\usepackage[utf8]{inputenc}
\usepackage{subfig}

\DeclareCaptionFormat{cont}{#1 (cont.)#2#3\par}

\newcommand{\halpha}{\ensuremath{\textrm{H}\alpha}\xspace}
\newcommand{\hanii}{\ensuremath{\textrm{H}\alpha+[\mathrm{N}\textsc{ ii}]}\xspace}
\newcommand{\hbeta}{\ensuremath{\textrm{H}\beta}\xspace}
\newcommand{\pabeta}{\ensuremath{\textrm{Pa}\beta}\xspace}
\newcommand{\paalpha}{\ensuremath{\textrm{Pa}\alpha}\xspace}
\newcommand{\OIII}{\ensuremath{[\mathrm{O}\textsc{ iii}]}\xspace}
\newcommand{\NII}{\ensuremath{[\mathrm{N}\textsc{ ii}]}\xspace}

\newcommand{\FeII}{\ensuremath{[\mathrm{Fe}\textsc{ ii}]}\xspace}
\newcommand{\HeI}{\ensuremath{[\mathrm{He}\textsc{ i}]}\xspace}

\newcommand{\AV}{\ensuremath{A_{\mathrm{V}}}\xspace}

\newcommand{\Flambda}
{\ensuremath{F_\lambda}\xspace}
\newcommand{\F}
{\ensuremath{F}\xspace}

\newcommand\ngals{14 }  % Keep a space after the number

\shorttitle{Emission Line Measurements from Medium-Band Photometry in MegaScience}
\shortauthors{Lorenz et al.}

\graphicspath{./}

\begin{document}

\title{Measuring Emission Lines with JWST-MegaScience Medium-Bands:\\A New Window into Dust and Star Formation at Cosmic Noon}

\author[0000-0002-5337-5856]{Brian Lorenz}
\affiliation{Department of Astronomy, University of California, Berkeley, CA 94720, USA}

\author[0000-0002-1714-1905]{Katherine A. Suess}
\affiliation{Department for Astrophysical and Planetary Science, University of Colorado, Boulder, CO 80309, USA}

\author[0000-0002-7613-9872]{Mariska Kriek}
\affiliation{Leiden Observatory, Leiden University, P.O. Box 9513, 2300 RA Leiden, The Netherlands}

\author[0000-0002-0108-4176]{Sedona H. Price}
\affiliation{Department of Physics and Astronomy and PITT PACC, University of Pittsburgh, Pittsburgh, PA 15260, USA}

\author[0000-0001-6755-1315]{Joel Leja}
\affiliation{Department of Astronomy \& Astrophysics, The Pennsylvania State University, University Park, PA 16802, USA}
\affiliation{Institute for Computational \& Data Sciences, The Pennsylvania State University, University Park, PA 16802, USA}
\affiliation{Institute for Gravitation and the Cosmos, The Pennsylvania State University, University Park, PA 16802, USA}

\author[0000-0002-7524-374X]{Erica Nelson}
\affiliation{Department for Astrophysical and Planetary Science, University of Colorado, Boulder, CO 80309, USA}

\author[0000-0002-7570-0824]{Hakim Atek}
\affiliation{Institut d’Astrophysique de Paris, CNRS, Sorbonne Universit\'e, 98bis Boulevard Arago, 75014, Paris, France}

\author[0000-0001-5063-8254]{Rachel Bezanson}
\affiliation{Department of Physics and Astronomy and PITT PACC, University of Pittsburgh, Pittsburgh, PA 15260, USA}

\author[0000-0001-6755-1315]{Gabriel Brammer}
\affiliation{Cosmic Dawn Center (DAWN), Copenhagen, Denmark}
\affiliation{Niels Bohr Institute, University of Copenhagen, Jagtvej 128, Copenhagen, Denmark}
\author[0000-0002-7031-2865]{Sam E. Cutler}
\affiliation{Department of Astronomy, University of Massachusetts, Amherst, MA 01003, USA}

\author[0000-0001-8460-1564]{Pratika Dayal}
\affiliation{Kapteyn Astronomical Institute, University of Groningen, 9700 AV Groningen, The Netherlands}

\author[0000-0002-2380-9801]{Anna de Graaff}
\affiliation{Max-Planck-Institut f\"ur Astronomie, K\"onigstuhl 17, D-69117, Heidelberg, Germany}

\author[0000-0002-5612-3427]{Jenny E. Greene}
\affiliation{Department of Astrophysical Sciences, Princeton University, 4 Ivy Ln., Princeton, NJ 08544, USA}

\author[0000-0001-6278-032X]{Lukas J. Furtak}
\affiliation{Physics Department, Ben-Gurion University of the Negev, P.O. Box 653, Be’er-Sheva 84105, Israel}

\author[0000-0002-2057-5376]{Ivo Labb\'e}
\affiliation{Centre for Astrophysics and Supercomputing, Swinburne University of Technology, Melbourne, VIC 3122, Australia}

\author[0000-0001-9002-3502]{Danilo Marchesini}
\affiliation{Department of Physics \& Astronomy, Tufts University, MA 02155, USA}

\author[0000-0003-0695-4414]{Michael V. Maseda}
\affiliation{Department of Astronomy, University of Wisconsin-Madison, 475 N. Charter St., Madison, WI 53706, USA}

\author[0000-0001-8367-6265]{Tim B. Miller}
\affiliation{Center for Interdisciplinary Exploration and Research in Astrophysics (CIERA), Northwestern University, 1800 Sherman Ave, Evanston IL 60201, USA}

\author[0000-0002-9816-9300]{Abby Mintz}
\affiliation{Department of Astrophysical Sciences, Princeton University, 4 Ivy Ln., Princeton, NJ 08544, USA}

\author[0000-0001-7300-9450]{Ikki Mitsuhashi}
\affiliation{Department for Astrophysical and Planetary Science, University of Colorado, Boulder, CO 80309, USA}

\author[0000-0002-9651-5716]{Richard Pan}
\affiliation{Department of Physics \& Astronomy, Tufts University, MA 02155, USA}

\author[0009-0001-0715-7209]{Natalia Porraz Barrera}
\affiliation{Department for Astrophysical and Planetary Science, University of Colorado, Boulder, CO 80309, USA}

\author[0000-0001-9269-5046]{Bingjie Wang (\begin{CJK*}{UTF8}{gbsn}王冰洁\ignorespacesafterend\end{CJK*})}
\affiliation{Department of Astronomy \& Astrophysics, The Pennsylvania State University, University Park, PA 16802, USA}
\affiliation{Institute for Computational \& Data Sciences, The Pennsylvania State University, University Park, PA 16802, USA}
\affiliation{Institute for Gravitation and the Cosmos, The Pennsylvania State University, University Park, PA 16802, USA}

\author[0000-0003-1614-196X]{John R. Weaver}
\affiliation{Department of Astronomy, University of Massachusetts, Amherst, MA 01003, USA}

\author[0000-0003-2919-7495]{Christina C.\ Williams}
\affiliation{NSF’s National Optical-Infrared Astronomy Research Laboratory, 950 North Cherry Avenue, Tucson, AZ 85719, USA}

\author[0000-0001-7160-3632]{Katherine E. Whitaker}
\affiliation{Department of Astronomy, University of Massachusetts, Amherst, MA 01003, USA}
\affiliation{Cosmic Dawn Center (DAWN), Denmark}

\begin{abstract}
We demonstrate the power of JWST-NIRCam medium-band photometry to measure emission line fluxes and study dust and star formation properties of galaxies at cosmic noon. In this work, we present photometric emission line measurements and spatially-resolved maps of \halpha and \pabeta for a sample of \ngals galaxies at $1.3\leq z\leq 2.4$, observed by the MegaScience medium-band survey and the UNCOVER deep spectroscopic survey. We measure line fluxes directly from the medium-band photometry and compare with spectroscopic measurements from UNCOVER. We find reasonable agreement between the photometric and spectroscopic emission line fluxes for both \halpha and \pabeta, with scatter $<0.15$ dex down to emission line equivalent widths of $10$\AA. We also make a nebular dust measurement from the ratio \pabeta / \halpha, finding an average nebular \AV of 1.4. Our photometric \AV measurements show a slightly larger scatter of $0.5$ magnitudes when compared to spectroscopic measurements; however, this scatter may be partially caused by aperture effects. Finally, we produce spatially resolved maps of \halpha emission, \pabeta emission, and stellar continuum. We find that offsets in \halpha and \pabeta emission are common, especially for galaxies with the highest \AV, indicating dusty sub-structures. Furthermore, the correlation between \halpha and continuum emission decreases with increasing \AV, suggesting that the dustiest objects have clumpy dust and star formation distributions. Our study demonstrates the power of medium-band photometry to directly probe emission line strengths, star formation, and dust attenuation for hundreds of galaxies in UNCOVER and thousands of galaxies in upcoming JWST medium-band surveys.
\end{abstract}


\section{Introduction} \label{sec:intro}

Understanding dust is essential to accurately understand observations of distant galaxies. Measurements of star formation rate (SFR), the shape of galaxy spectral energy distributions (SEDs), and the stellar mass growth of galaxies over time are all significantly influenced by dust attenuation \citep{calzetti_dust_2001}. In order to better interpret observations, we must understand both the total quantity and spatial distribution of dust in distant galaxies. 

However, dust at cosmic noon is still not fully understood. The amount of dust in galaxies tends to increase with increasing stellar mass and SFR \citep[e.g.,][]{garn_predicting_2010, dominguez_dust_2013, price_direct_2014, whitaker_constant_2017, cullen_vandels_2018, battisti_average_2022, shapley_mosfire_2022, runco_mosdef_2022, maheson_unravelling_2024}, but may not be controlled by these properties alone. The shape of the dust attenuation curve is still uncertain, as well as which factors influence this shape \citep[e.g.,][]{kriek_dust_2013, reddy_mosdef_2015, reddy_mosdef_2020, salmon_breaking_2016, salim_dust_2020, shivaei_mosdef_2020}. Additionally, $z=2$ star-forming galaxies tend to have messy, clumpy morphologies \citep{elmegreen_resolved_2007, swinbank_hubble_2010, tadaki_nature_2014}, and so the geometry of dust in these systems is not well understood. Dusty, submillimeter-detected galaxies at cosmic noon have equally complex dust geometries \citep{gillman_structure_2024}, with visible dust lanes that seem to trace the NIR stellar continuum \citep{price_uncover_2025}. A recent study of dusty star-forming galaxies measured their attenuation curves, finding that they have lower $R_V$ than expected at bluer wavelengths \citep{cooper_rubies_2025}. Previous studies indicate that dust may be clumped in small spherically symmetric regions rather than spread throughout the ISM \citep{reddy_mosdef_2015, lorenz_updated_2023, lorenz_stacking_2024}, and may also exist in large, kiloparsec-scale star-forming clumps \citep{schreiber_constraints_2011, wuyts_smoother_2012}. Finally, studies have probed the geometry of dust and star formation at cosmic noon with spatially resolved maps \citep{nelson_bd_2016, tacchella_dust_2018, matharu_first_2023} suggesting higher dust-obscured star formation in the centers, but the full picture of where dust and star formation occur within these distant galaxies remains uncertain.

One of the best ways to examine dust is through spectroscopic ratios of hydrogen emission lines, as their intrinsic ratios are well-understood assuming case B recombination, so any deviations from these ratios are caused by dust \citep{reddy_paschen-line_2023}. Unfortunately, spectroscopy of galaxies is typically observationally expensive, requiring long observations of a limited number of objects in a field. Alternatively, dust can be studied through stellar population synthesis modeling of galaxy photometry \citep[e.g. \texttt{Prospector},][]{leja_deriving_2017, johnson_stellar_2021}, but many assumptions go into modeling galaxy SEDs, including the age-dust-metallicity degeneracy \citep{papovich_stellar_2001}, causing larger uncertainties for their dust contents. It is also possible to study dust through its infrared emission \citep{boquien_new-generation_2021}, but for most galaxies, this emission is difficult to observe. These observational challenges have made it difficult to collect large samples of high-quality dust measurements at cosmic noon. 

Fortunately, with the launch of the James Webb Space Telescope \citep[JWST,][]{gardner_james_2023}, observers now have access to another method to directly study dust quantity and geometry: medium-band photometry with NIRCam \citep{rieke_performance_2023}. The JWST medium-bands have typical widths of 0.2$\mu$m, which are narrow enough to contain only one or two strong emission lines in their bands at specific redshifts. Additionally, the full suite of NIRCam medium bands densely samples wavelength space, allowing for a local continuum measurement to be made near each strong emission line. Instead of relying on spectroscopy, observers are now able to measure emission line fluxes directly from photometry \citep[e.g.,][]{geach_hizels_2008, roberts-borsani_improving_2021, simmonds_ionizing_2023, withers_spectroscopy_2023}, allowing for larger-scale surveys of dust attenuation at cosmic noon.  Additionally, the infrared capabilities of JWST allow us to push these emission line measurements to even higher redshifts, reaching out to cosmic noon. Since they are such a powerful tool, a number of medium-band surveys have already been conducted (e.g., JEMS, \citealt{williams_jems_2023}; CANUCS/Technicolor, \citealt{willott_canucs_2023}; JOF, \citealt{eisenstein_jades_2023}; MegaScience, \citealt{suess_medium_2024}), with more planned in Cycle 4 (e.g.PID 7814 MINERVA and PID 8559 SPAM)

In this work, we demonstrate a method for measuring \halpha and \pabeta emission line fluxes from photometry alone at $1.3<z<2.4$, the redshift range that contains both emission lines in the medium-bands. This sample is a pilot study that contains \ngals galaxies that have both medium-band photometry from MegaScience \citep{suess_medium_2024} and spectroscopy from UNCOVER \citep{bezanson_jwst_2024, price_uncover_2024} to assess the accuracy of this technique, similar to \citet{withers_spectroscopy_2023}. We show that the photometric emission line fluxes are in good agreement with the spectroscopy. In combination with the low photometric errors, this implies that this technique is a promising way to conduct much larger surveys of SFR and dust at cosmic noon. Additionally, we use the same technique for measuring line fluxes on the medium-band images, creating spatially resolved maps of \halpha and \pabeta that can be used to trace dust and star formation.

Throughout this work we assume a $\Lambda$CDM cosmology with $\Omega_m=0.3$, $\Omega_{\Lambda}=0.7$, and $H_0=70$ km s$^{-1}$ Mpc$^{-1}$. 

\section{Data and Sample Selection} \label{sec:data}

We obtain our sample of galaxies with NIRCam medium-band photometry and NIRSpec prism spectroscopy from the UNCOVER and MegaScience surveys \citep{bezanson_jwst_2024, suess_medium_2024}. The UNCOVER survey targets a field covering the Abell 2744 cluster, a gravitational lens with a very wide area of high magnification. This field was additionally selected due to its overlap with the Hubble Frontier Fields \citep{lotz_frontier_2017}. The MegaScience survey complements UNCOVER, filling in the medium-bands that had not yet been observed. Together, these surveys cover the field with all 20 broad- and medium-band NIRCam filters, providing a powerful dataset for this work. With it, we can capture \halpha and \pabeta emission lines in the medium bands at a wide range of redshifts. 

We use the public DR3 photometric catalog \citep[][internal catalog v5.2.0]{suess_medium_2024} and the associated \texttt{Prospector} fits  \citep[][v5.3.0]{wang_uncover_2024}. The \texttt{Prospector} fits give photometric redshifts, stellar mass, SFR, metallicity, and dust measurements. The $\approx$70,000 sources are fit with \texttt{Prospector}-$\beta$ priors on mass and star formation history, and the parameter space is explored with \texttt{dynesty} \citep{speagle_dynesty_2020}. Of particular interest to this work, the dust is fit with a two-component model \citep{charlot_simple_2000} --- one component attenuating all light, and an additional component that affects only young stars --- with a variable-slope \citet{noll_analysis_2009} attenuation curve. 

Additionally, we use the public DR4 NIRSpec prism spectra from UNCOVER \citep{jakobsen_near-infrared_2022, boker_-orbit_2023, price_uncover_2024}. These low-resolution spectra allow for accurate redshift measurements, and a direct measurement of emission line fluxes. We use the spectra to make direct emission line measurements (Section \ref{subsec:flux_cal}) as a comparison sample to assess the accuracy of photometric emission line measurements. In addition, these accurate redshifts allow us to locate exactly where emission lines land in the photometric filters, as well as identify any other contaminating lines in the medium-bands.

\begin{figure*}[]
\vglue -5pt
\includegraphics[width=\textwidth]{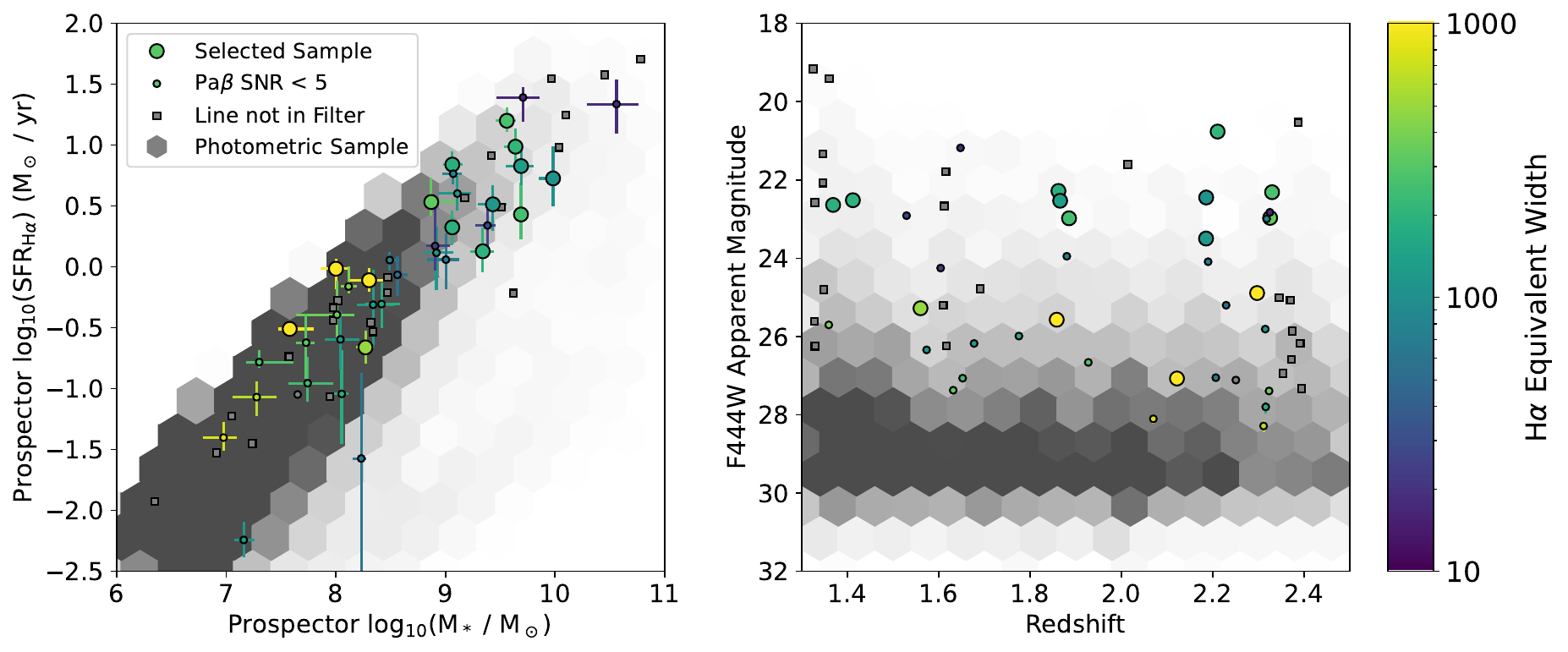}
\caption{
Our selected sample of \ngals galaxies (large circles) shown with the full UNCOVER sample (gray hexagons) in stellar mass vs. SFR (measured by \texttt{Prospector}) and F444W apparent magnitude vs redshift. Points are color-coded by \halpha equivalent width as measured fromt he prism spectra. Squares show galaxies where the spectroscopic redshift indicates that one of the emission lines is not fully contained in the medium-band photometry. Small circles show objects that do not have high enough \pabeta signal-to-noise ratios (SNR$< 2$) to make accurate emission line measurements. We see that our sample includes galaxies on both sides of the star-forming main sequence. Galaxies are spread across the $1.3\leq z\leq 2.4$ range and detected down to 27th magnitude in F444W. 
}
\label{fig:sample_select}
\end{figure*}

We aim to select a sample of galaxies for which we can make both photometric and spectroscopic measurements of the \halpha and \pabeta emission lines, allowing us to measure nebular \AV. We impose the following selection criteria onto the spectroscopic sample of nearly 700 galaxies. First, we ensure high-quality spectra by selecting objects with \texttt{flag\_zspec\_qual} = 3, following \citet{price_uncover_2024}. Next, we place a redshift cut of $1.3\leq z\leq 2.4$ to ensure that both the \halpha and \pabeta emission lines are observed in one of the medium bands and that there is a continuum band on both sides of the emission line. We then discard any galaxies where the filter transmission is $\leq 50\%$ of its maximum within 20\AA\ of the spectroscopic center of the emission line, as these are possibly too close to the filter edge. Finally, we require that both the spectroscopically measured \halpha and \pabeta line fluxes have a signal-to-noise ratio greater than 2. We note that more than half of the possible targets are excluded due to low \pabeta signal-to-noise ratios, so we could double the sample size for science cases that only require \halpha. The selected sample of \ngals galaxies is shown in the context of the full photometric sample in Figure \ref{fig:sample_select}. The galaxies are distributed equally above and below the star-forming main sequence within $\approx$0.5 dex. They are spread throughout the redshift range $1.3\leq z\leq 2.4$ with apparent F444W magnitudes less than 27 and most with $\log(M_*/M_\odot)>9$. Our sample consists of preferentially higher-mass galaxies that have typical SFRs for their mass range.

\section{Measuring Emission Lines from Photometry and Spectroscopy} \label{sec:data_analysis}

\subsection{Measuring Emission Lines from Medium-Band Photometry} \label{subsec:emission_line_seds}

We measure the fluxes of the \halpha and \pabeta emission lines from only three medium-band observations per line, which we will compare to the spectroscopic fluxes. The process is nearly identical for \halpha and \pabeta, with the exception of a few correction factors discussed below. 

First, we identify the continuum filters as the medium bands surrounding the target emission line. For example, if \pabeta falls within the F360M band, then F335M and F410M are used for continuum measurement. The wavelength coverage of some of the medium bands overlap \citep{rieke_performance_2023}. If an emission line falls in multiple bands, we use the next-closest filter for continuum measurement. 

To compute  the continuum flux $F_{\lambda\mathrm{,cont}}$, we linearly interpolate the flux measurement between the two continuum filters to the effective wavelength of the target filter. Then, we subtract $F_{\lambda\mathrm{,cont}}$ from the flux density in the target filter. The remaining flux density in the target filter should be entirely due to the line. Next, we convert the excess flux density due to the line to a total flux by multiplying by the effective width of the target filter;

\begin{equation}
    F_\mathrm{line} = \Flambda \times \mathrm{width}.
\end{equation}
These width values are roughly 2000\AA\ for \halpha filters and 4000\AA\ for \pabeta filters \citep{rieke_performance_2023}. We do not apply any corrections for the varying response function across the line profile since the filters have nearly flat response functions, which would lead to less than a 10\% change in flux at worst. We have ensured through selection that the lines do not fall near the edges of the filters, and therefore should have strong transmission. As we expand this technique to purely photometric samples, uncertain redshifts will make correcting for filter transmission very challenging. To ensure that this pilot study provides a fair representation of the errors, we choose not to correct for this $<10\%$ effect in our spectroscopic subsample. Fortunately, we find a low median redshift uncertainty in MegaScience of 0.035 by comparing the spectroscopic redshift to the photometric redshifts for galaxies in our targeted redshift range \citep{price_uncover_2024, wang_uncover_2024}. This redshift uncertainty corresponds to roughly 250\AA\ at the wavelength of \halpha at $z=2$. The typical filter width is 2000\AA, so even in fully photometric samples, we can determine that the line is contained within the filter to high degrees of certainty.

Unfortunately, neither \halpha nor \pabeta can be completely isolated in their photometric filters. The \halpha line is blended with \NII at 6550\AA\ and 6585\AA\ in both the UNCOVER prism spectra and medium-band photometry, while \pabeta is typically blended with \FeII 12570\AA\ in the photometry, but well-resolved in the prism spectra. We correct for each of these cases.

The ratio of \NII/\halpha is well-studied at cosmic noon, and depends strongly on the galaxy's metallicity. We estimate the \NII6585\AA/\halpha ratio using the fundamental metallicity relation (FMR) from \citet{sanders_mosdef_2021} for MOSDEF galaxies, which is a redshift-independent link between mass, SFR, and gas-phase metallicity. For our sample, we compute a gas-phase metallicity from the FMR using the \texttt{Prospector} mass and SFR measurements, then an \NII6585\AA/\halpha value from that metallicity, finding a median ratio of 0.09. Finally, we take the theoretical transition probability ratio of \NII6550\AA\ to \NII6585\AA\ as 1/3 \citep[see][table 1]{dojcinovic_flux_2023}. We use these estimates to correct both the spectroscopic and photometric measurements of \halpha to remove the contribution of the \NII doublet. 

We only need to correct for \FeII in the photometric measurements for \pabeta. The \FeII line seems to be associated with shocked gas \citep{hartigan_infrared_2004}, and therefore may be indicative of AGN \citep{calabro_near-infrared_2023}. Unlike the \NII line, there is not a strong relation for \FeII with galaxy properties, so we use the objects in our sample with spectroscopic \FeII detections to extrapolate the correction factor for objects without spectroscopic \FeII. To determine the factor for this correction, we fit the flux of the \FeII emission lines with signal to noise ratios of at least 2 for all of the objects with UNCOVER spectra at $1.3\leq z\leq 2.4$. The \FeII fluxes for these 10 objects do not show strong correlations with redshift, \texttt{Prospector} mass, metallicity, or SFR. We correct each object where \FeII is detected at 2$\sigma$ with its individual measurement, and correct by the median \FeII/\pabeta value (0.27) for the other objects. Additionally, \HeI 10830\AA\ occasionally falls within the blue continuum band for \pabeta. Correcting for this emission line spectroscopically only changes the result by at most 2\%, so we opt to not make any corrections for \HeI. 

Both \halpha and \pabeta emission lines have underlying stellar absorption. These absorption lines affect both the photometry and spectroscopy equally, but the absorption line strength differs for \halpha and \pabeta. Thus, the absorption lines do not impact our comparison of the spectroscopic and photometric fluxes from the same line, but slightly impact the line ratio \pabeta /\halpha. Therefore, we make a correction for hydrogen absorption lines when computing the line ratio. For each galaxy, we measure the equivalent widths of its absorption features from its \texttt{Prospector} stellar population synthesis model without emission lines, and use these absorption line strengths to correct both the photometric and spectroscopic line fluxes.

We measure uncertainties for all objects with Monte Carlo simulations, perturbing the three photometric data points randomly by their uncertainties and recalculating the flux. We carry these simulations through all the way to an \AV measurement, taking 1$\sigma$ uncertainties as the 16th and 84th percentiles of 1000 draws.

We apply the same method of line flux measurement to the full images (PSF-matched to F444W) to measure emission spatially. Figure 3 of \citet{suess_medium_2024} shows that the PSF-matching is reliable to sub-1\%. Pixel by pixel, we repeat the above method using the medium-band images, but do not apply the \NII correction. We generate these maps for both \hanii and \pabeta, and propagate uncertainties from the associated noise maps. The \pabeta line is fairly weak, so most objects do not have high signal-to-noise measurements of \pabeta in these images. When visualizing the line maps, we show only \pabeta contours around the regions with high signal-to-noise ratios. 

\begin{figure*}[tp]
\vglue -5pt
\includegraphics[width=\textwidth]{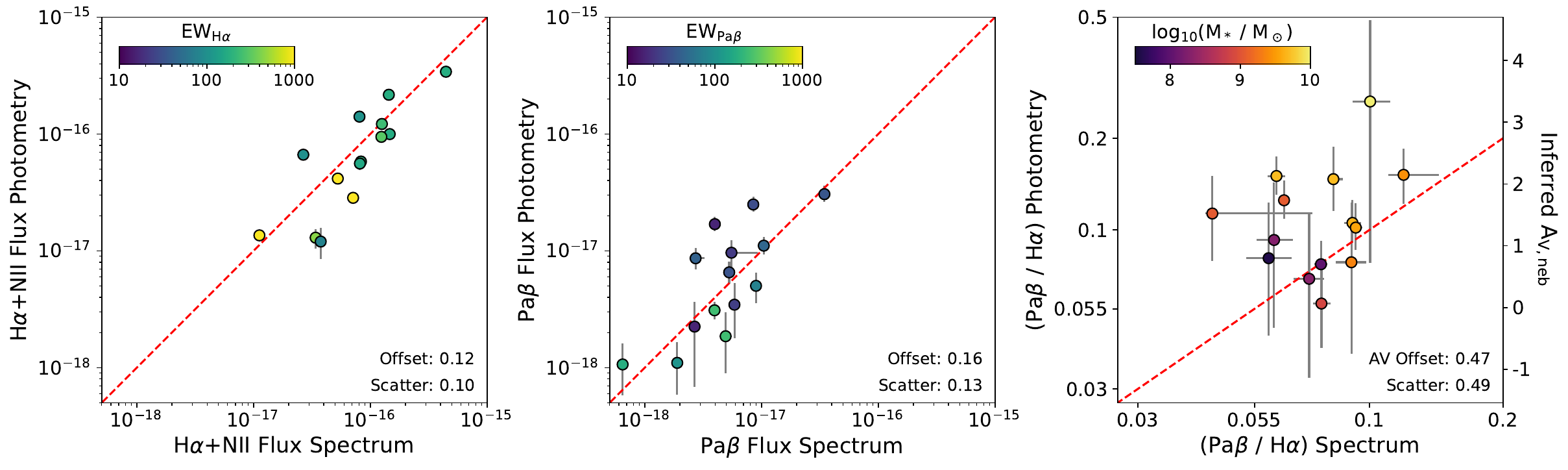}
\caption{
Comparisons between photometric (y-axis) and spectroscopic (x-axis) measurement of \halpha line flux, \pabeta line flux, and $\frac{\pabeta}{\halpha}$ ratio for the \ngals galaxies in our sample. Points are color-coded by photometric equivalent width in the first two panels and \texttt{Prospector} mass in the right panel. Inferred \AV measurements are shown on the right axis of the line ratio plot (right panel). Median offset and 1$\sigma$ scatter are shown at the bottom right of each plot. Uncertainties are computed from Monte Carlo simulations. We see strong agreement with a scatter of less than 0.15 dex in both lines, demonstrating that the medium bands are a powerful tool for measuring isolated emission lines.}
\label{fig:flux_compare}
\end{figure*}

\subsection{Flux Calibration and Emission Fitting} \label{subsec:flux_cal}

In order to assess the accuracy of photometric emission line fits, we also measure the \halpha and \pabeta line fluxes spectroscopically. We first flux-calibrate the UNCOVER prism spectroscopy by fitting a fifth-order polynomial to the 20-band JWST photometry, finding correction factors ranging from 1.3 to 6.0. We then fit each of the emission lines directly from the flux-calibrated spectra. We linearly fit the continuum around each of the emission lines from 5500-7700\AA\ and 11800-14200\AA, masking the regions from 6200-7000\AA\ and 12200-13100\AA\ to avoid contamination from the lines themselves. After continuum subtracting the lines, we simultaneously fit \halpha and \pabeta with Gaussian profiles, allowing the amplitude and width of each line to vary. We find that the blended \halpha + \NII complex is not perfectly centered at a rest wavelength of 6563\AA, so we allow for a redshift offset between \halpha and \pabeta of $z_\mathrm{offset}\leq0.005$. This accounts for both the varying \NII line width and the uncertain NIRSpec wavelength calibration \citep{deugenio_jades_2024, graaff_rubies_2025}. We estimate uncertainties on the fit fluxes and line ratios with 1000 Monte Carlo simulations.

\subsection{\AV measurement} \label{subsec:make_dust_maps}

With both photometric and spectroscopic line flux measurements, we can compute a nebular \AV. The intrinsic ratio of \halpha to \pabeta is 18:1 assuming case B recombination \citep{reddy_paschen-line_2023}. Any deviation from this ratio is due to dust affecting the bluer \halpha line more strongly than the redder \pabeta. 

We compute the \AV nebular values:
\begin{equation}
    \AV = R_V \times k \times \log_{10}\left(\frac{\halpha / \pabeta}{18}\right),
\end{equation}
where $k$ is computed as
\begin{equation}
    k = \frac{2.5}{k'\mathrm{(\halpha)} - k'\mathrm{(\pabeta})},
\end{equation}
and $k'(\lambda)$ is described in \citet{calzetti_dust_2000}. We assume $R_V = 4.05$.

\section{Results}
\label{sec:results}

In this paper, we demonstrate the use of NIRCam medium-band photometry to measure nebular emission lines, dust properties, and create emission line maps. To this end, we measure \halpha and \pabeta emission lines for \ngals galaxies from both medium-band photometry and NIRSpec prism spectroscopy. First, we use the spectra to assess the accuracy of photometric emission line flux and nebular \AV measurements. Then, we present spatially resolved maps of both \halpha and \pabeta in each galaxy, allowing us to study the locations of dust and star formation.

\subsection{Accuracy of Emission Line Measurements}
\label{subsec:SED_dust}

In Figure \ref{fig:flux_compare}, we compare the photometric and spectroscopic line flux measurements and \AV. We find strong agreement in \hanii and \pabeta line fluxes with scatters of 0.10 and 0.13 dex, respectively. Even at low \pabeta signal-to-noise ratios, the individual flux measurements are still fairly accurate. While we are directly measuring the line flux from the photometric bands, we note that the spectroscopic \halpha equivalent widths range from 100\AA\ to 1000\AA\ (typically 200\AA), and the \pabeta equivalent widths range from 10\AA\ to 250\AA\ (typically 40\AA).

Similarly, the spectroscopic and photometric line ratios \pabeta/\halpha are reasonably consistent, with a scatter of 0.49 in \AV. Uncertainties are fairly large, and primarily driven by the uncertainty in \pabeta measurement. With the points color-coded by mass, we generally recover the well-known relation that more massive galaxies are more dusty. 

For many points, the measurement uncertainties do not reach the one-to-one line, particularly in the individual emission line flux comparisons. This result points towards other systematics dominating the uncertainties of this technique, rather than measurement errors. One likely systematic uncertainty may be due to slit losses between the photometry and spectroscopy. For example, a slit that is offset from regions of high star-formation (i.e. object 44283) could cause a difference in the \halpha flux measured by spectroscopy and photometry. We attempted to improve the accuracy of the nebular \AV measurement by redoing the calculation with only the pixels contained in the MSA shutters, but this correction does not reduce the observed scatter. However, we noted that five out of the six farthest points from the one-to-one line show clearly visible offsets between the shutter locations and the peak of either the \hanii or \pabeta emission line map. Additionally, the six objects closest to the one-to-one line do not have clear visual offsets between the shutter location and the peak of the emission line map. These tests point towards some differences in coverage between the shutters and aperture photometry as a source of error.  

\citet{withers_spectroscopy_2023} conducted a similar study, measuring the equivalent widths of \halpha and \OIII+\hbeta from medium-band photometry and NIRSpec prism spectroscopy for 15 objects at $1.7\leq z\leq 6.7$. They seem to find a slightly tighter relation for the \halpha flux, and similar scatter to our work for \OIII+\hbeta. The reasonable consistency of our measurements points towards the medium-bands as an excellent means to measure strong, isolated emission line fluxes photometrically. These methods allow us to measure bright emission line fluxes for large samples of galaxies without requiring spectra.

\subsection{Emission Line and Dust Maps at Cosmic Noon}
\label{subsec:dust_results}

With the same technique used to calculate photometric emission line fluxes, we created \halpha and \pabeta line maps (Figure \ref{fig:dust_map_mosaic0}). The leftmost column shows a three-color image, with \hanii falling in the green filter, and the position of the NIRSpec shutters shown in white. The second column shows \halpha contours overlaid on the continuum map (measured near \halpha). The third column shows the \hanii linemap. Finally, due to low signal-to-noise ratios of the \pabeta maps, we show the \pabeta contours overlaid on the \halpha maps in the fourth column. 

The majority of the objects (9/14) seem to have their \halpha emission overlapping with their brightest points in the continuum at the galaxy's center. These overlaps may be indications of strong central star-formation, as found by other studies at cosmic noon \citep[e.g.,][]{barro_sub-kiloparsec_2016, ellison_star_2018}. However, as the integrated \AV increases, the \halpha emission tends to be visually offset from the peak continuum flux, as seen in all three of the most dusty galaxies. Notably, object 54625 shows a very strong secondary clump of \halpha emission that does not show up strongly in the continuum. We quantified the offsets between \halpha and continuum emission by measuring the Pearson correlation coefficient ($r$) between every pixel in the emission line map and continuum map within the galaxy. We show these $r$ values plotted against our medium-band \AV measurement in Figure \ref{fig:pearson_r}. We find that the correlation coefficient decreases as \AV increases, indicating that dustier galaxies tend to have larger differences in their \halpha and continuum maps. This quantitatively confirms the trends that we can see visually. These offsets between \halpha and continuum emission may indicate the presence of dust or galaxy growth, as the galaxies are forming stars in regions where there are not many stars already. 

The galaxies in our sample have SFRs that range from 0.1-50 $\mathrm{M}_\odot / $yr, and therefore show strong \halpha emission. The majority of these galaxies show large regions of \halpha emission that either cover most of the galaxy or are grouped into multiple clumps. Prior observations at cosmic noon find star formation occurring in large, kpc-scale clumps \citep{schreiber_constraints_2011, wuyts_smoother_2012}. Our maps are qualitatively consistent with these findings, showing large regions of strong \halpha emission. The aforementioned object 54625 may be the clearest example of one such large star-forming clump, with object 44283 potentially showing another. 

By examining the strong regions of \pabeta emission compared to the \halpha maps, we find that \pabeta appears to be generally slightly offset from the \halpha emission (9/14 galaxies). Regions where we observe \halpha but not \pabeta are likely due to the low SNR of the \pabeta maps. On the other hand, locations where \pabeta is visible without \halpha are likely indicating the presence of dust, which suppresses \halpha flux much more strongly than \pabeta. All 5 of the galaxies with photometric \AV $> 1.5$ show these offsets. The offsets tend to be fairly small ($<0.1$''), with the strongest region of \pabeta being close to the strongest region of \halpha. This proximity may indicate that the dust is somewhat patchy, with some star-forming regions being obscured, while others are not. Dusty star-forming regions are consistent with prior studies which claim that dust at cosmic noon is clumpy and found around star-forming regions rather than evenly spread throughout the ISM \citep{reddy_mosdef_2015, lorenz_updated_2023, lorenz_stacking_2024}. Patchy star-formation is also consistent with the fact that \halpha and continuum emission are more offset as \AV increases, as the dust is not spread evenly among the star-forming regions. Finally, the centers of the dustiest objects show \pabeta emission, indicating that the centers may have higher dust obscuration.

One of our objects, 37182, shows a very strong signal in \pabeta to the lower right of the galaxy that is not seen in the \halpha emission or continuum maps. This indicates an extremely dusty region. The nature of this source is unclear --- perhaps this is a galaxy merger with a second core that is very dust-obscured. 

\begin{figure*}[tp]
\vglue -5pt
\includegraphics[width=\textwidth]{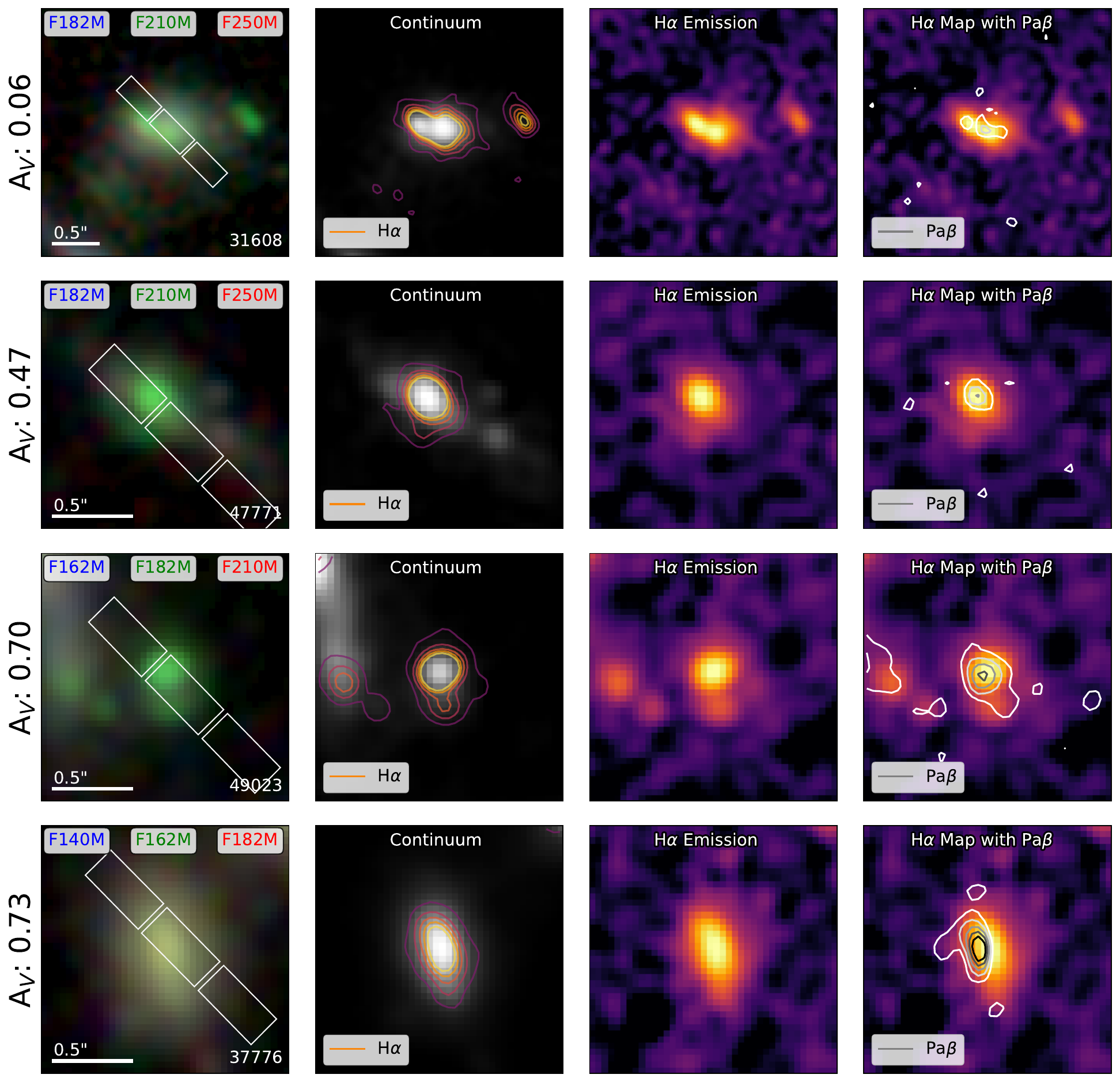}
\caption{
Images and line maps for each of our \ngals objects, sorted by measured photometric \AV. First column: RGB image, with the \halpha emission line falling in the green medium-band, and the position of the NIRSpec slit shown in white. We show a 0.5'' scale bar, as well the DR3 ID from the UNCOVER public catalogs. Second column: continuum map (measured near \halpha) overlaid with \halpha emission map contours at signal-to-noise ratios of 1 through 5. Third column: \halpha emission line map. Fourth column: The same \halpha emission line map overlaid with \pabeta map contours at signal-to-noise ratios of 1 through 5. Locations with strong \pabeta but low \halpha are likely dusty.}
\label{fig:dust_map_mosaic0}
\end{figure*}

\begin{figure*}[tp]
\vglue -5pt
\ContinuedFloat
\captionsetup{list=off,format=cont}
\centering
\includegraphics[width=\textwidth]{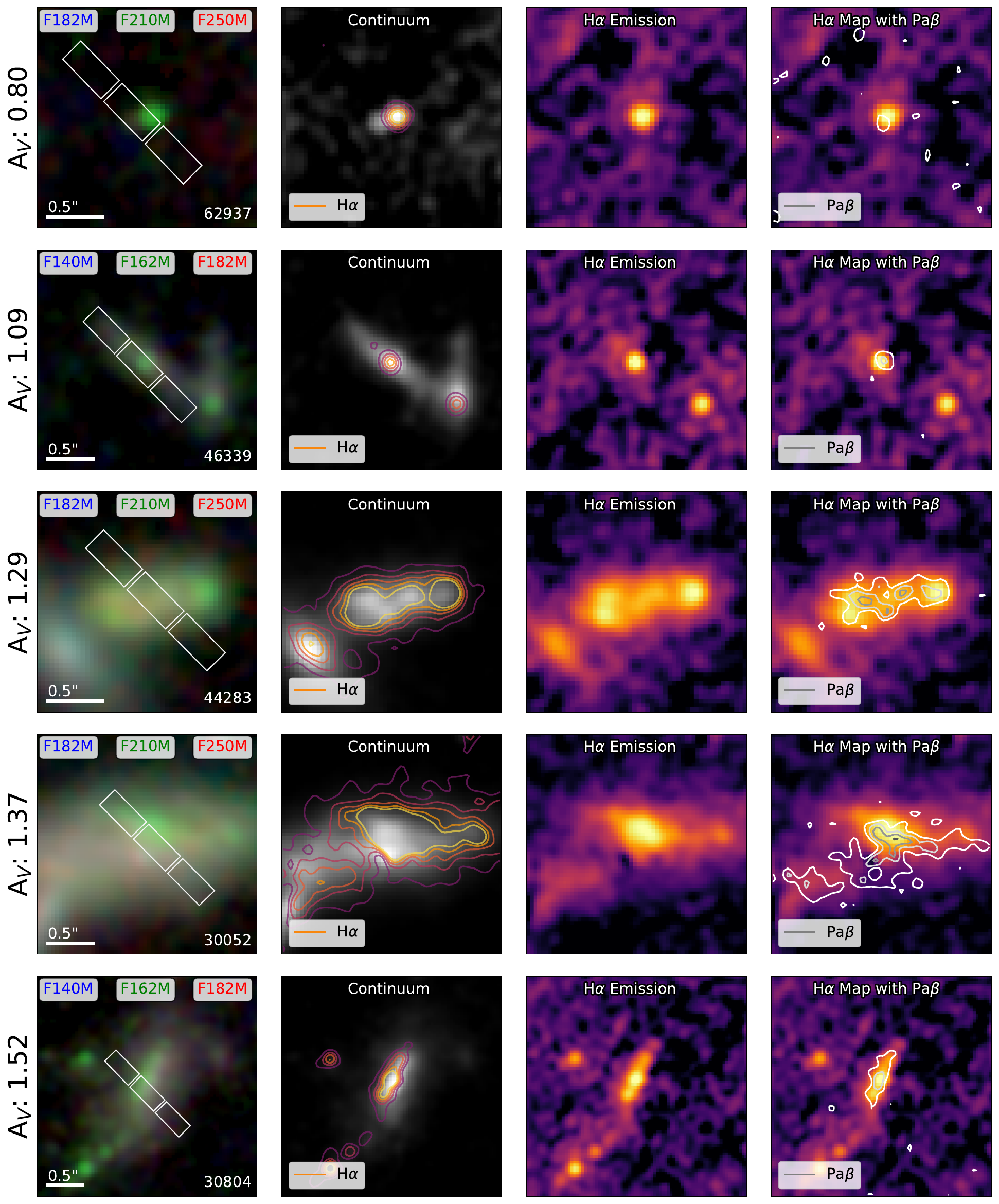}
\caption{
}
\label{fig:dust_map_mosaic1}
\end{figure*}

\begin{figure*}[tp]
\vglue -5pt
\ContinuedFloat
\captionsetup{list=off,format=cont}
\centering
\includegraphics[width=\textwidth]{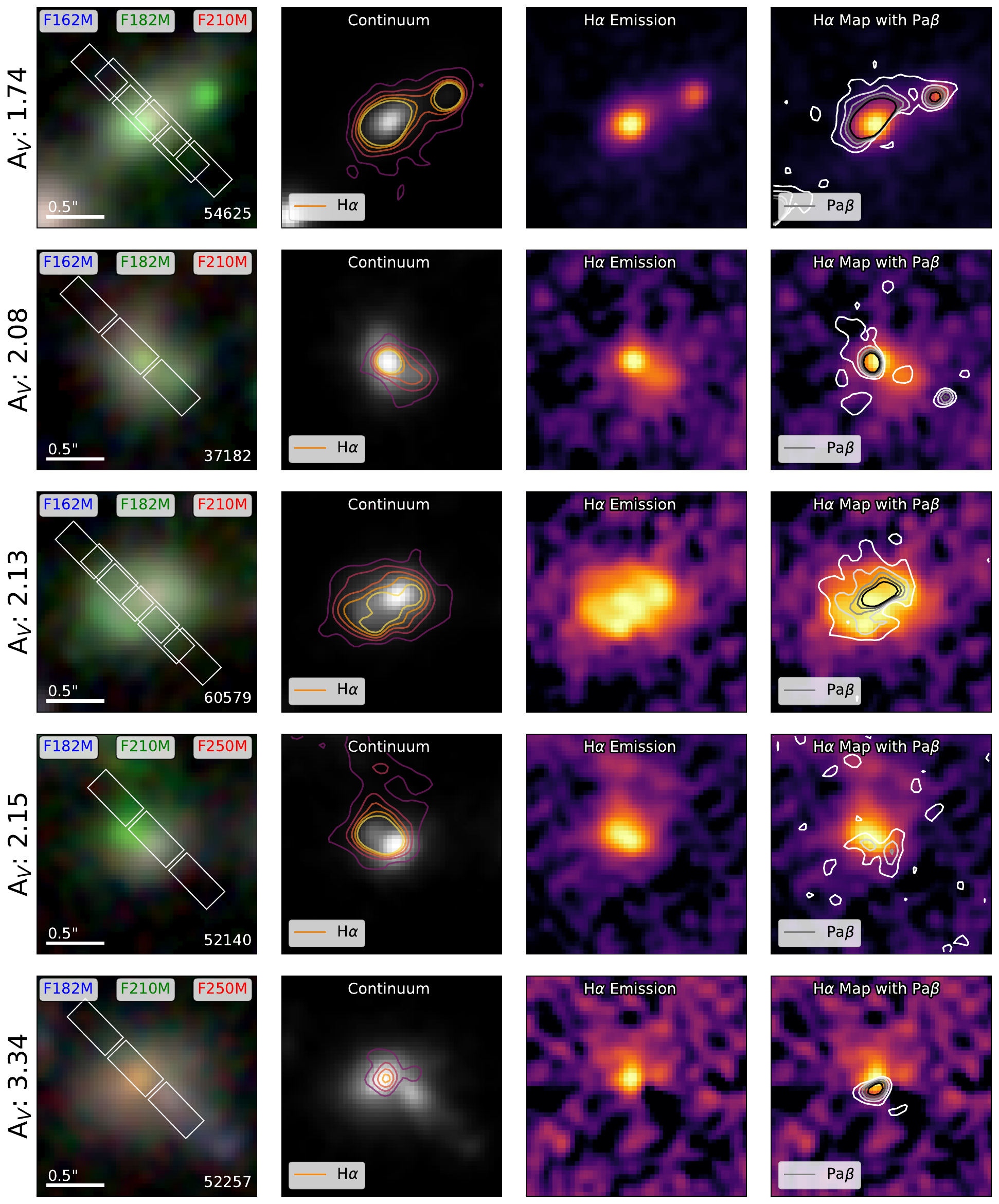}
\caption{
}
\label{fig:dust_map_mosaic2}
\end{figure*}

\begin{figure}[tp]
\vglue -5pt
\includegraphics[width=0.5\textwidth]{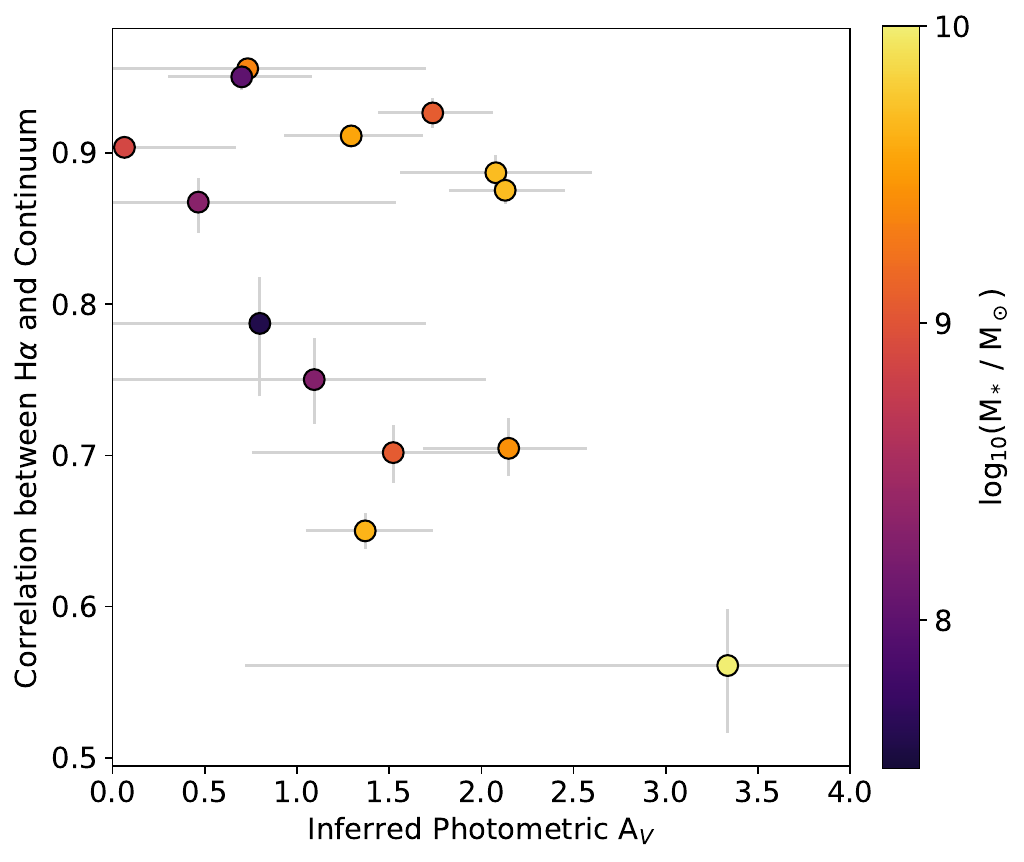}
\caption{We show the Pearson correlation coefficient ($r$) computed between the \halpha emission line map (see Figure \ref{fig:dust_map_mosaic0}) and continuum map. We only include pixels that were flagged as part of the galaxy through its segmentation map from UNCOVER. $r$ tends to decrease as photometric \AV increases, indicating an offset between the brightest pixels in \halpha and the continuum. The offsets suggest that dust is obscuring some of the star-forming regions, but not others, indicating patchy dust distribution. They may also hold information about galaxy growth, as new star are forming in regions where there are not many older stellar populations.}
\label{fig:pearson_r}
\end{figure}

\section{Discussion and Conclusions} \label{sec:summary}

In this work, we measured both the total flux and spatial distribution of \halpha and \pabeta emission lines from the MegaScience NIRCam medium-band photometry for a sample of \ngals galaxies at cosmic noon, and then compared these measurements to NIRSpec prism spectra from the UNCOVER survey. We summarize our main findings:

\begin{itemize}
    \item The photometric and spectroscopic measurements of the \halpha and \pabeta line flux are generally in agreement with a scatter of $<0.15$ dex (Figure \ref{fig:flux_compare}), giving confidence to the accuracy of measuring isolated, high equivalent-width line fluxes from medium-band photometry. 
    \item We measured the nebular \AV from the $\frac{\pabeta}{\halpha}$ ratio, finding reasonable agreement between the photometry and spectra with a scatter of 0.5 dex. The photometric nebular \AV increases with increasing galaxy mass, as expected.
    \item We created spatially resolved \halpha emission line maps and overlaid \pabeta line map contours on them. There appear to be offsets between rest-frame optical stellar continuum, \halpha emission, and \pabeta emission, particularly in the most dusty systems. Our findings are consistent with studies suggesting patchy dust geometries \citep{reddy_mosdef_2015, lorenz_updated_2023, lorenz_stacking_2024}.
\end{itemize} 

Our study demonstrates that the medium bands are capable of measuring isolated emission lines with a scatter of $0.15$ dex down to equivalent widths of 10\AA. However, applying this technique to a sample that has only medium-band photometry will bring its own set of challenges. Since galaxies would not have spectroscopically confirmed redshifts, it will be difficult to ensure that the targeted line falls entirely within the photometric filter. However, we find that the median redshift uncertainty in MegaScience is 0.035 for galaxies in our targeted redshift range. This uncertainty is on the order of 250\AA\ at the wavelength of \halpha, so the typical redshift uncertainty is much smaller than the typical filter width of 2000\AA. Therefore, redshift uncertainties likely will not be a concern for the majority of star-forming galaxies with \halpha emission.

Applying these photometric emission line measurements to larger numbers of galaxies can unlock new ways to study dust and star formation at cosmic noon. In the current UNCOVER field, this would allow for emission line measurements and maps for hundreds of galaxies without requiring the observationally expensive spectra. While we restricted our sample to \halpha and \pabeta, other strong emission lines, such as \OIII+\hbeta or \paalpha, should be measurable. With this technique, we can conduct large surveys of star formation rate or dust properties without requiring spectra, which currently is limited to less than 1000 galaxies across the full redshift range for UNCOVER.  

The medium bands also allow us to generate spatially resolved emission line maps for a much larger sample of galaxies, giving insight into the locations of dust and star formation at a wide range of redshifts. The maps have spatial resolutions of $\approx$1kpc at $z=2$ and show clear structure even with per-pixel signal-to-noise ratios of 1, allowing for detailed studies of where star formation is occurring, including whether galaxies are showing inside-out or outside-in growth \citep[e.g.,][]{nelson_where_2016, tacchella_dust_2018}. Additionally, lensing from the Abell 2744 cluster allows for even higher spatial resolution for some galaxies in the full UNCOVER/MegaScience photometric sample. 

This pilot study shows very promising potential for the JWST medium-bands. They hold the ability to effectively isolate individual emission lines, measure their fluxes, and spatially resolve their locations with maps. With the wealth of data available through recent surveys \citep[e.g.,][]{williams_jems_2023, willott_canucs_2023, eisenstein_jades_2023, suess_medium_2024}, and more medium-band surveys on the horizon (e.g., JWST 7814 MINERVA), the medium-bands provide an exciting new lens into galaxies at cosmic noon. 

\begin{acknowledgments}
This work is based on observations made with the NASA/ESA/CSA James Webb Space Telescope. The raw data were obtained from the Mikulski Archive for Space Telescopes at the Space Telescope Science Institute, which is operated by the Association of Universities for Research in Astronomy, Inc., under NASA contract NAS 5-03127 for \textit{JWST}. These observations are associated with JWST Cycle 1 GO program \#2651 and JWST Cycle 2 GO program \#4111, and this project has gratefully made use of a large number of public JWST programs in the Abell 2744 field including JWST-GO-2641, JWST-ERS-1324, JWST-DD-2756, JWST-GO-2883, JWST-GO-3538, and JWST-GO-3516. Support for program JWST-GO-4111 was provided by NASA through a grant from the Space Telescope Science Institute, which is operated by the Associations of Universities for Research in Astronomy, Incorporated, under NASA contract NAS5-26555.
\end{acknowledgments}

\typeout{}\bibliography{MegaScience}{}
\bibliographystyle{aasjournal}

\end{document}